\documentstyle[epsfig,aps,preprint]{revtex}
\tightenlines
\begin{document}
\draft
\preprint{USC-PHYS(NT)-99-03}
\title{A next-to-next-to-leading-order $pp \rightarrow pp\pi^0$ \\
Transition Operator in Chiral Perturbation Theory}
\author{V. Dmitra\v{s}inovi\'{c}$^a$, K. Kubodera$^a$, 
F. Myhrer$^a$ and T. Sato,$^b$ }
\address{$^a$ Department of
Physics and Astronomy,
University of South Carolina \\
Columbia, SC 29208, U. S. A.} 
\address{$^b$Department of Physics, Osaka University,
Toyonaka, Osaka 560-0043, Japan}

\date{\today}

\maketitle
\vspace{-0.5cm}
\begin{abstract}
We present a systematic analysis of next-to-next-to-leading-order
diagrams that contribute to the $pp \rightarrow pp\pi^0$ production 
at threshold. 
Analytic expressions are given for the effective transition operators,
and the relative importance of various types of diagrams is discussed.
The vertex-correction-type graphs are found to give only small corrections 
to lower order graphs in conformity with expectations.
By contrast, we find very large contributions
from two-pion exchange graphs that can be interpreted as a part of
effective $\sigma$-meson exchange diagrams.
%%%%%%% ADDED %%%%%%%%%%
The recoil correction to the pion rescattering diagram also turns out to be 
large.
%%%%%%% ADDED %%%%%%%%%%
\end{abstract}
\pacs{PACS numbers: 25.40.Ep, 25.40.Qa, 12.39.Fe}

\narrowtext

\paragraph{Introduction}
High-precision measurements\cite{meyetal90,Uppsala} of 
neutral pion production in proton-proton collisions 
$pp \rightarrow pp\pi^0$ just above the threshold
have generated renewed theoretical scrutiny 
of this reaction \cite{ms91}-\cite{kmr96}.
This reaction is unique among two-nucleon pion production processes
in that it is not well described by the single-nucleon process 
(Born term), Fig.1(a), and the
s-wave pion rescattering process, Fig.1(b). 
The reason is that the ``large" Weinberg-Tomozawa term 
does not contribute to the $pp \rightarrow pp\pi^0$ reaction, in contrast 
with e.g. charged-pion production $pp \rightarrow pn\pi^+$,
thus rendering $pp \rightarrow pp\pi^0$ particularly sensitive to 
and hence 
an interesting testing ground for the less-well-understood ``small" 
isoscalar s-wave pion rescattering terms.
From the calculations done thus far it is  clear that 
the two most basic processes, Fig.1(a) and 1(b), 
give much smaller $pp \rightarrow pp\pi^0$
cross sections than the measured values.
Lee and Riska's model calculation \cite{lr93} suggests
that shorter-range isoscalar meson-exchange processes, 
like $\sigma$- and $\omega$- exchanges between the 
two protons, might be very important in this reaction. 

In heavy-baryon chiral perturbation theory [HB$\chi$PT] 
\cite{jm91} one can define the 
``large'' and ``small'' terms in the Hamiltonian
by way of chiral-order counting.
%\footnote{
In this language the first pion rescattering contributions to 
$pp \rightarrow pp\pi^0$ are one chiral-order higher than 
those of the charged-pion production 
(e.g., $p p \rightarrow p n \pi^+$). 
This means that $pp \to pp\pi^0$ production is 
very sensitive to ``correction terms", 
which tend to be ``masked" by the leading terms 
in most low-energy pion-nucleon processes. 
Since the Born term and the pion rescattering term 
do not explain the $pp \to pp\pi^0$ cross section, we extend here
our previous calculation \cite{pmmmk96} to the next chiral order. 

It must be mentioned, however, 
that application of HB$\chi$PT 
to the $NN\rightarrow NN\pi$ processes 
is a delicate matter in at least two aspects.
First, the non-negligible energy-momentum transfers involved
can make less clear the distinction between 
reducible and irreducible diagrams 
in Weinberg's chiral counting scheme.
This has led Cohen {\it et al.} \cite{cfmv96} to propose
new counting rules to be used for inelastic processes
like $NN\rightarrow NN\pi$.
It is noteworthy that in this modified scheme
loop diagrams can be of the lowest order. 
Secondly, as the nucleon recoil involved becomes appreciable,
the static nucleon ``propagator'', 
which is one of the basic features of the HB$\chi$PT formalism, 
can become increasingly problematic. 
Meanwhile, ``improving" the HB$\chi$PT propagator
by including a nucleon recoil term 
%%
%%               INSERTED               %%%%%%%%%%
requires an extension of HB$\chi$PT from its original form.
%%                END of INSERTED LINE    %%%%%%
%%
%%               DELETED                 %%%%%%%%%%
%% is not strictly within the confines of HB$\chi$PT
%% because changing the nucleon propagator
%% implies abandoning the very concept of chiral order counting. 
%%              END OF DELETED LINES    %%%%%%%%%%
%%
Despite the importance of these issues,
we postpone the discussion of these matters to a future publication.
In this Letter we rather use the standard counting rules
\`{a} la Weinberg\cite{wei90}
within the framework of HB$\chi$PT.
We concentrate on calculations of the effective transition operators 
strictly within the narrow definition of HB$\chi$PT.
We also relegate to the future 
the consideration of the initial- and final-state interactions 
in the transition amplitude,
even though we are aware of the paramount importance
of a full distorted-wave (DW) calculation 
for comparison with the experiment.
This is, for one thing, because
the transition operators resulting from 
our systematic treatment of the new diagrams
exhibit very complicated energy- and momentum-dependencies,
rendering a full DW calculation a highly non-trivial numerical task. 
Secondly, we wish to separate out the effects of higher-order diagrams 
in the ``kernel'' (irreducible part) of the reaction from the perhaps more
mundane yet numerically important initial- and final-state interaction
effects.

The principal purpose of this note is 
%two-fold: (a) 
to report analytic results for the transition operators
that arise from a systematic treatment 
of the next chiral order diagrams in HB$\chi$PT. 
In order to 
%have an estimate of 
gain some insight into 
the relative numerical significance
of these diagrams, we compare their absolute values, at the threshold, 
to that of the one-pion exchange rescattering graph, Fig.1(b).
Our most important finding 
%of our present study
is that some of the two-pion exchange diagrams, 
which have no lower-order counterparts, 
give by far the dominant transition operators at the threshold
(sometimes by an order of magnitude larger than 
the nominally ``leading'' ones).
These new diagrams 
%can perhaps 
may be interpreted as a part of an effective 
$\sigma$-meson exchange (see below).
%our results, albeit subject to modification 
%due to DW effects, give the first hint
%of a connection between ChPT and the phenomenological one-boson 
%exchange models.

\paragraph{Conventions and other preliminaries}
The effective Lagrangian ${\cal L}_{\rm{ch}}$
in HB$\chi$PT is expanded as 
\begin{equation}
{\cal L}_{\rm{ch}} = {\cal L}^{(0)} + {\cal L}^{(1)}
+ {\cal L}^{(2)}+ \cdots\;\; .
\label{eq:Lag0}
\end{equation}
${\cal L}^{(\bar{\nu})}$ represents a term of chiral order $\bar{\nu}$
with $\bar{\nu}\equiv d + (n/2) - 2$, where $n$ is the number of
fermion fields in the term, and $d$ is the number of derivatives or
powers of $m_\pi$.  The explicit forms for the $\bar{\nu}=0$ and $ 1 $ 
terms are \cite{bkm95}: 
\begin{eqnarray}
{\cal L}^{(0)} &=&
 \frac{f^2_\pi}{4} \,\mbox{Tr}
[ \partial_\mu U^\dagger \partial^\mu U
 + m_\pi^2 (U^\dagger +  U - 2) ]
 + \bar{N} ( i v \!\cdot\! D + g_A^{} S \!\cdot\! u ) N
\label{eq:L0}\\
{\cal L}^{(1)} &=&
-\frac{i g_A^{}}{2m_N} \bar{N}
\{ S\!\cdot\!D, \,v \!\cdot\! u \} N
 + 2c_1 m_\pi^2 \bar{N} N \mbox{Tr}
( U + U^\dagger - 2 )  \nonumber \\
 &&+ (c_2 \!-\! \frac{g_A^2}{8m_N}) \bar{N}
(v \!\cdot\! u)^2 N
 + c_3 \bar{N} u \!\cdot\! u  N + \cdots 
\label{eq:L1}
\end{eqnarray} 
where we have retained only terms
of direct relevance to our present calculation. 
The SU(2) field matrix $U(x)$ is 
non-linearly related to the pion field
and has standard chiral transformation properties.
We use the representation
$U(x) = \sqrt{1-[\bbox{\pi}(x)/f_\pi]^2}
+i\bbox{\tau} \!\cdot\! \bbox{\pi}(x)/f_\pi$
as in Ref. \cite{bkm95}.
$N(x)$ represents the large component 
of the heavy-nucleon field;
the four-velocity parameter $v_\mu$
is chosen to be $v_\mu=(1,0,0,0)$;
%%%%%%%%% THE SIGN CHANGED %%%%%%%%%%
$D_\mu N = \left(\partial_{\mu} + 
{1 \over 2} \left[ \xi^{\dagger}, \partial_{\mu}
\xi \right] \right)\! N$ is the covariant derivative of $N$;
$S_\mu$ is the (covariant) spin operator, 
which in the nucleon rest frame
%for our choice of $v_\mu$ 
becomes $S^\mu$ = $(0, \vec{\sigma}/2)$, and 
$u_\mu = i [\xi^\dagger \partial_\mu \xi - 
\xi \partial_\mu \xi^\dagger]$, 
where
$\xi = \sqrt{U(x)}$ \cite{bkm95}.
The low-energy constants $c_1,c_2$ and $c_3$ 
have been determined from other processes, see e.g.
Refs. \cite{slmk97,bkm95}. 
An explicit expression for ${\cal L}^{(2)}$, which includes 
${\cal O }(m_N^{-2})$ recoil terms 
as well as terms containing new low-energy constants,
can be found, e.g., in Ref. \cite{fms98}.

In Weinberg's chiral counting \cite{wei90}
each irreducible Feynman diagram carries a 
chiral order index $\nu$ defined by
$\nu = 4 - E_N - 2C + 2L + \sum_i \bar{\nu}_i$,
where $E_N$ is the number of nucleons
in the Feynman diagram, $L$ the number of loops,
$C$ the number of disconnected parts
of the diagram, and the sum runs over all the vertices
in the Feynman graph \cite{wei90}.
In this note we consider irreducible diagrams with chiral orders
up to $\nu$ = 2 that give rise to $pp \rightarrow pp \pi^0$ 
transition operators.
These diagrams are shown in Figs. 1-5. 
In the following ${\cal T}^{(\nu)}$ stands for a transition operator 
of chiral order $\nu$.
The lowest-order transition operator for 
the Born diagram [Fig.(1a)]
has $\nu=-1$, and that for the 
rescattering diagram has $\nu=+1$ [Fig.(1b)]. 
We define the first HB$\chi$PT calculation \cite{pmmmk96,cfmv96} 
which includes the aforementioned terms as 
the next-to-leading (NLO) 
calculation. Hence we decree our next chiral order 
($\nu$ = 2) calculation to be  
next-to-next-to-leading (NNLO) order. 
With the use of ${\cal L}_{\rm{ch}}$ in Eq.(\ref{eq:Lag0}), 
the two transition operators 
that feature in the NLO calculation 
are given (in momentum space)
by \cite{pmmmk96,cfmv96}
\begin{eqnarray}
{\cal T}^{Born} = {\cal T}^{(\nu = -1)}
& = &
\frac{g_A}{2f_\pi}
\sum_{j=1,2} 
%\left( -\vec{\sigma}_j\cdot\vec{q} +
\frac{\omega_q}{2m_N}\vec{\sigma}_j
\cdot (\vec{p}_j + \vec{p}_j^{\,\,\prime} )
%\right) 
\tau_j^0,
\label{eq:Tminus1} \\
{\cal T}^{Resc} = {\cal T}^{(\nu = +1)}
&=& 
\left( \frac{- g_A}{f_\pi}\right) 
\sum_{j=1,2} \kappa(k_j,q)\frac{\vec{\sigma}_j
\cdot \vec{k}_j\tau_j^0}{k_j^2 - m_\pi^2 +i\eta},
\label{eq:Tplus1}
\end{eqnarray}
where $\vec{p}_j$ and $\vec{p}^{\,\,\prime}_j$ ($j=1,2$)
denote the initial and final
momenta of the $j$-th proton.
The four-momentum of the exchanged pion is defined by the
nucleon four-momenta at the $\pi NN$ vertex:
$k_j \equiv p_j-p^{\,\,\prime}_j$, where 
$p_j =(E_{p_j}, \vec{p}_j),
p^{\,\,\prime}_j=(E_{p^{\,\,\prime}_j}, 
\vec{p}^{\,\,\prime}_j)$ 
with the definition 
$E_p=(\vec{p}^{\,\,2} + m_N^2)^{1/2} - m_N$.
The s-wave rescattering vertex function
$\kappa(k,q)$ is calculated from Eq.(\ref{eq:L1}):
\begin{equation}
\kappa(k,q)\equiv \frac{m_\pi^2}{f_\pi^2}\,
\left( \,2c_1 - (c_2 - \frac{g_A^2}{8m_N})
\frac{\omega_q k_0}{m_\pi^2}
 - c_3 \frac{q\cdot k}{m_\pi^2}\,\right)\;,
\label{eq:kappakq}
\end{equation}
where $k=(k_0,\vec{k})$ and $q=(\omega_q, \vec{q})$
represent the four-momenta of the exchanged and final pions,
respectively, and $\omega_q=(\vec{q}^{\; 2} +m_\pi^2)^{1/2}$.

\paragraph{Analytic results}
When we consider NNLO, 
we find 19 topologically distinct new types of diagrams 
that can potentially contribute to $NN \rightarrow NN\pi$ reactions. 
For a particular case of the $pp \rightarrow pp\pi^0$ reaction near threshold,
the isospin selection rules and the s-wave character of the outgoing pion
reduce this number from 19 to 6.
We refer to these six as Type I, Type II, $\ldots$, Type VI,
and illustrate them in Figs. 2-4.
Types I, II and III appear in Fig. 2, Type IV in Fig. 3,
and Types V and VI in Fig.4. 
Some of these diagrams have been considered 
by Gedalin  {\it et al.}\cite{gmr98}, as we shall discuss later.
We denote by $M_{\mbox{\scriptsize I}}$, 
$M_{ \mbox{\scriptsize II}}$, ..., 
the NNLO transition operators 
that arise from the diagrams of Type I, II,..., respectively.
In addition to these six operators, as will be explained in more detail below,
there is an NNLO contribution from Fig. 5, 
which looks the same as Fig. 1(b), but 
which constitutes a higher-order correction to 
${\cal T}^{(+1)}$ [eq.(\ref{eq:Tplus1})].
We denote by $M_{\mbox{\scriptsize VII} }$
the transition operator due to this correction.
Then the total NNLO transition operator we consider is given by 
\begin{equation} 
{\cal T}^{(+2)} = M_{\mbox{\scriptsize I}} +
 M_{\mbox{\scriptsize II}} + M_{\mbox{\scriptsize III}} 
+ M_{\mbox{\scriptsize IV}} + M_{\mbox{\scriptsize V}}
+M_{\mbox{\scriptsize VI}} +M_{\mbox{\scriptsize VII}}.
\label{eq:Tplus2}
\end{equation} 
The explicit expressions for these operators are as follows.
(In the expressions below
the subscripts $i$ and $j$ refer to nucleon number 1 and 2.) 
\begin{eqnarray}
M_{\mbox{\scriptsize I} } & = &
\frac{g_A}{8f_{\pi}^5} \sum_{i\ne j=1,2} 
S_i \!\cdot\!k_j\,[ X_1 + \frac{X_2}{\vec{k}_j^2}] 
\label{eq:2a}
\end{eqnarray} 
where 
\begin{eqnarray} 
X_1 & = & v \!\cdot\! (p_i\!+\!p_i')\; I_{\pi}(k_j) 
+ v \!\cdot\! (q\!+\!\frac{ p_i'}{2}) \; 
v\!\cdot\! (q\!+\!p_i\!+\!p_i')
\,I_0(v\!\cdot\! p_i ,\!-\! v\!\cdot\! k_j , k_j^2)  + 
\nonumber \\
& + & 
v\!\cdot\! (q\!-\!\frac{ p_i}{2} ) \; 
v\!\cdot\! (q\!-\!p_i\!-\!p_i') 
\; I_0(v\!\cdot\! p_i', v\!\cdot\! k_j , k_j^2)  
\label{eq:2a1} \\ 
X_2 & = & 
%\nonumber \\  &-& 
- v\!\cdot\! (p_i\!+\!p_i') \; v\!\cdot\! (5q\!-\!k_i) \; 
v\!\cdot\! k_j \; 
I_{\pi}(k_j) - 
v\!\cdot\! (q\!+\!\frac{p_i'}{2}) v\!\cdot\! (q\!+\!p_i\!+\!p_i') 
\times 
\nonumber \\
& \times & 
{}[J_0(v\!\cdot\! p_i) \!-\! 
J_0(v\!\cdot\! (p_i\!+\!k_j)) 
\!+\! v\!\cdot\! k_j v\!\cdot\! (k_j\!+\!2p_i ) 
I_0(v\!\cdot\! p_i,\!-\! v\!\cdot\! k_j, k_j^2 )] + \nonumber \\
  & -& v\!\cdot\! (q\!-\! \frac{p_i}{2} ) 
v\!\cdot\! (q\!-\!p_i\!-\!p_i') 
{}[J_0(v\!\cdot\! p_i') \!-\!J_0(v\!\cdot\! (p_i'\!-\!k_j)) + 
v\!\cdot\! k_j v\!\cdot\! (k_j \!-\!2p_i')
I_0(v\!\cdot\! p_i', v\!\cdot\! k_j , k_j^2 ) ]  . 
\label{eq:2a2} 
\end{eqnarray}
\begin{eqnarray}
M_{\mbox{\scriptsize II}} & = &
      \frac{g_A^3}{8f_{\pi}^5} \sum_{i\ne j=1,2} 
S_i \!\cdot\! (k_j \!-\! k_i )
 \{ J_0(v\!\cdot\! p_j)
\!+\!J_0(v\!\cdot\! p_j')\!+\! v\!\cdot\! (p_j\!+\!p_j') 
\; I_{\pi}(k_j) + \nonumber \\
 & & +  (\!-\!2m_\pi^2\!+\!2v\!\cdot\! p_j \; 
v\!\cdot\! p_j'\!+\!k_j^2) 
\; I_0(v\!\cdot\! p_j',\!-\! v\!\cdot\! k_j, k_j^2) \} . 
\label{eq:2b}
\end{eqnarray}
\begin{eqnarray}
M_{\mbox{\scriptsize III}} & = &
      \frac{g_A^3}{4f_{\pi}^5} \sum_{i\ne j =1,2} [
  S_i\!\cdot\! k_j \bar{X}_1 + 2
 i\epsilon_{\mu\nu\alpha\beta}v^{\alpha}S_j^{\beta}
S_i^{\mu}k_j^{\nu}\bar{X}_2]  , 
\label{eq:2c} 
\end{eqnarray} 
where 
\begin{eqnarray} 
\bar{X}_1 
& = & -v\!\cdot\! (p_i\!+\!p_j\!+\!p_i'\!+\!p_j') 
 \frac{1}{2} I_\pi(k_j) + \nonumber \\
 & & + \frac{1}{v\!\cdot\! (p_i\!-\!p_j')}[
      v\!\cdot\!(p_i\!+\!q\!+\!k_j){\cal Y}
(v\!\cdot\! p_i,\!-\!k_j) 
    \!-\! v\!\cdot\!(p_j'\!+\!q\!+\!k_j){\cal Y}
(v\!\cdot\! p_j',\!-\!\!k_j)] + \nonumber \\
 & & + \frac{1}{v\!\cdot\! (p_i'\!-\!p_j)}[
      v\!\cdot\!(p_i'\!-\!q\!-\!k_j){\cal Y}
(v\!\cdot\! p_i',k_j) 
    \!-\!v\!\cdot\!(p_j\!-\!q\!-\!k_j){\cal Y}
(v\!\cdot\! p_j,k_j)] 
\label{eq:2c1} \\ 
\bar{X}_2 & = & 
\frac{1}{v\!\cdot\! 
(p_i\!-\!p_j')}[v\!\cdot\!(p_i\!+\!q\!+\!k_j)
{\cal A}(v\!\cdot\! p_i,\!-\!k_j)
 \!-\!v\!\cdot\!(p_j'\!+\!q\!+\!k_j){\cal A} 
(v\!\cdot\! p_j',\!-\!k_j)] + 
 \nonumber \\
 &  & -\frac{1}
{v\!\cdot\! (p_i'\!-\!p_j)}[v\!\cdot\!(p_i'\!-\!q\!-\!k_j) 
{\cal A}(v\!\cdot\! p_i',k_j)
  \!-\!v\!\cdot\!(p_j \!-\!q\!-\!k_j){\cal A}(v\!\cdot\! p_j,k_j)]\,.
\label{eq:2c2} 
\end{eqnarray}
In Eqs.(\ref{eq:2c1}) and (\ref{eq:2c2}): 
\begin{eqnarray}
\lefteqn{{\cal Y}(\omega,P)  = } \nonumber\\
& &\frac{1}{4}[4 (v\!\cdot\! P) I_{\pi}(P)\!-\!
J_0(\omega)\!-\!J_0(\omega\!-\! v\!\cdot\! P)
\!+\! (2m_\pi^2\!-\!2(\omega\!-\!v\!\cdot\! P)^2 \!+\! \vec{P}^2) 
\, I_0(\omega,v\!\cdot\! P, P^2) ] + \nonumber \\
& &
+ \; \left( \frac{[2m_\pi^2\!-\!\omega^2\!-\!(\omega\!-\!v\!\cdot\! P)^2]}
{4\vec{P}^2}\right) \times 
\nonumber \\ 
& & \times [ 
   2 (v\!\cdot\! P) I_{\pi}(P) \!-\!J_0(\omega)
\!+\!J_0(\omega\!-\!v\!\cdot\! P)
\!+\! ( v\!\cdot\! P)  (2\omega\!-\!v\!\cdot\! P) \, I_0(\omega,v\!\cdot\! P, 
P^2) ]  , 
\end{eqnarray}
and 
\begin{eqnarray}
{\cal A} (\omega, P) & = & 
\frac{(2\omega \!-\!v\!\cdot\! P) P^2}{4 \vec{P}^2} 
I_\pi (P) + \nonumber \\ 
& & + \left( \frac{m_\pi^2}{2} \!+\! 
\frac{P^2 [P^2 \!+\!4\omega^2\!-\! 
4\omega(v\!\cdot\! P)]}{8 \vec{P}^2} \right) 
I_0(\omega, v\!\cdot\! P,P^2) \!+\! \nonumber \\ 
& & -\frac{1}{4} 
J_0(\omega \!-\! v\!\cdot\! P) \!+\! 
\left( \frac{P^2 \!-\!2\omega (v\!\cdot\! P)}
{8 \vec{P}^2} \right) 
\left( J_0(\omega) \!-\! 
J_0(\omega \!-\! v\!\cdot\! P) \right)+ \nonumber \\ 
& & +\frac{v\!\cdot\! P \!-\! 2\omega}
{2 (4\pi)^2} \!+\! {\cal O}(4\!-\!d)  . 
\end{eqnarray} 
\begin{eqnarray}
M_{\mbox{\scriptsize IV}} & = &
      \frac{g_A^3}{4f_{\pi}^5} \sum_{i\ne j = 1,2} 
\frac{S_j\!\cdot\! k_j}
{k_j^2 \!-\! m_\pi^2 \!+\! i \eta}  \nonumber \\
 & & \times \{ \left(\!-\!m_\pi^2 \!+\! 
(v\!\cdot\! p_i)^2 \right) J_0(v\!\cdot\! p_i) 
   \!+\! \left(\!-\!m_\pi^2 \!+\! (v\!\cdot\! p_i')^2 \right) 
J_0(v\!\cdot\! p_i') + \nonumber \\ 
   & &   + v\!\cdot\! ( p_i \!+\! p_i' ) \Delta_\pi 
\!+\! [ 3 k_i^2 \!-\! m_\pi^2 \!-\! 2k_i\!\cdot\! q]X \} , 
\label{eq:3} 
\end{eqnarray} 
where 
\begin{eqnarray}
X & = & 
       -  \frac{v\!\cdot\! (p_i \!+\! p_i')}{2} \; 
I_\pi (k_i) + \nonumber \\ 
 & & + \left( m_\pi^2 \!-\! 
v\!\cdot\! p_i \; v\!\cdot\! p_i' \!-\! \frac{1}{2} k_i^2 
\right) 
\; I_0(v\!\cdot\! p_i, v\!\cdot\! k_i, k_i^2)
  \!-\!   \frac{1}{2} [ J_0(v\!\cdot\! p_i) \!+\! 
J_0(v\!\cdot\! p_i') ]  . 
\label{eq:3x} 
\end{eqnarray}
\begin{eqnarray}
M_{\mbox{\scriptsize V}}  & = & 
      \frac{g_A^3}{4f_{\pi}^5} \sum_{i\ne j =1,2} 
\frac{\!-\!S_j\!\cdot\! k_j}
{k_j^2\!-\!m_\pi^2\!+\!i\eta} \nonumber \\ 
& & \times {} \{ v\!\cdot\! (p_i\!+\!p_i')\Delta_{\pi}
    \!+\!((v\!\cdot\! p_i)^2\!-\!m_\pi^2)J_0(v\!\cdot\! p_i)
    \!+\!((v\!\cdot\! p_i')^2\!-\!m_\pi^2)J_0(v\!\cdot\! p_i') \} . 
\label{eq:4a} 
\end{eqnarray}
\begin{eqnarray}
M_{\mbox{\scriptsize VI}} & = &
      \frac{g_A}{8f_{\pi}^5} \sum_{i\ne j =1,2} 
\frac{\!-\!S_j\!\cdot\! k_j}{k_j^2\!-\!m_\pi^2\!+\!i\eta} \nonumber \\ 
 & & \times \{ v\!\cdot\! (p_i\!+\!p_i')\Delta_{\pi} \!+\! 
{} v\!\cdot\! (2q\!+\!p_i') \; v\!\cdot\! (\!-\! k_i\!+\!2q \!+\!p_i')\;
J_0(v\!\cdot\! (p_i'\!+\!q )) 
+ \nonumber \\
{}& & +
   v\!\cdot\!(2q\!-\!p_i ) \; v\!\cdot\! 
(\!-\! k_i\!+\!2q \!-\!p_i )\;J_0(v\!\cdot\! (p_i\!-\!q) ) \} . 
\label{eq:4b} 
\end{eqnarray}
As mentioned earlier, the graph in Fig.5 
with the relevant lowest order 
 vertices 
gives ${\cal T}^{(+1)}$ [eq.(\ref{eq:Tplus1})].
If we, however, use for the 
pion rescattering vertex in Fig.5 the 
lagrangian ${\cal L}^{(2)}$\cite{fms98},
there results an NNLO transition operator, $M_{\mbox{\scriptsize VII}}$,
which represents ``recoil" term corrections of ${\cal O}(m_N^{-2})$ 
to ${\cal T}^{(+1)}$.
Its explicit expression is
\begin{eqnarray} 
M_{\mbox{\scriptsize VII}} = \left( - \frac{g_A}{f_\pi}\right) 
\sum_{i\ne j =1,2} 
\kappa_i^\prime (k_j,q) 
\frac{\vec{\sigma}_j \!\cdot\! \vec{k}_j  \tau^0_j}
{k_j^2 - m_\pi^2 +i \eta } 
\label{eq:1b}
\end{eqnarray}
where the expression for $\kappa_i^\prime (k_j,q)$ 
is obtainable from Eq.(C.3) of Ref.\cite{fms98}. 

The above expressions for the $\nu$ = 2 diagrams
contain four independent one-loop integrals,
 $\Delta_\pi$, $J_0(\omega)$, $I_{\pi}(P^2 )$
and $I_0(\omega , v\!\cdot\! P, P^2 )$, of which 
the first three contain divergences.  
The finite parts are defined to include 
 $\ln(m_\pi /\lambda)$. We choose the cut-off 
 parameter to be 
$\lambda$ = 1 GeV.   
The  two integrals, $\Delta_\pi$ and  $J_0(\omega)$, 
can be found in Ref.\cite{bkm95}, 
while $I_{\pi}(P^2 )$ is a standard Feynman integral: 
\begin{equation}
I_{\pi}(P^2)  = 
\frac{1}{i}\int \frac{d^4 l}{(2\pi)^4}
\frac{1}{\left[m^2-l^2 - i\varepsilon \right]
\left[m^2-(l-P)^2 - i\varepsilon \right]} . 
\label{e:ipi} \\
\end{equation} 
The last integral, which is new, is given by 
\begin{eqnarray}
I_0 (\omega, v\!\cdot\! P , P^2) & = &
\frac{1}{i}\int \frac{d^4 l}{(2\pi)^4}
\frac{1}{( v\!\cdot\! l - \omega -i\varepsilon)
(m_\pi^2 - l^2 - i\varepsilon)
\left[m_\pi^2-(l-P)^2 - i\varepsilon \right]}
\nonumber \\
& = &
 \frac{1}{16\pi^2}\int_0^1 dx \left[
 \theta(s)\frac{2}{\sqrt{s}}
\left(\frac{\pi}{2} + \tan^{-1}\frac{\xi}{\sqrt{s}} \right)
+ \theta(-s)\frac{1}{\sqrt{-s}}\log \left|\frac{\xi-\sqrt{-s}}{\xi+\sqrt{-s}}
\right| \right] + 
 \nonumber \\
&+& i
\frac{1}{16\pi}\int_0^1 dx \frac{\theta(-s)}{\sqrt{-s}}
\left[\theta(\xi + \sqrt{-s}) + \theta(\xi - \sqrt{-s})\right],
\label{e:io} 
\end{eqnarray} 
where 
$s =  m^2- \omega^2 + x (2\omega P_0 - P^2) + x^2(P^2-P_0^2)$,  
$P_0 = v\!\cdot \! P$ and $\xi  = \omega - x P_0 $.
%for $P^2 < 0 $. 
$ I_0 (\omega, v\!\cdot\! P , P^2)$
reduces to the integral $\gamma_0(v\!\cdot\! P)$
given in appendix B of Ref.\cite{bkm95} in 
the limit $\omega$ = 0 and $P^2$ = 0.
This limit, however, is not applicable to the 
$pp \rightarrow pp\pi^0$ reaction. 

Our NNLO diagrams contain the usual divergences 
which need to be regularized and then renormalized 
by appropriate counter terms. 
The single-nucleon process (Born term) of Fig.1(a) 
receives two types of corrections: (i) the finite
``recoil corrections" of ${\cal O} (m_N^{-2})$ involving 
finite known parameters, and (ii) 
the (infinite) loop corrections.
The $\nu$ = 1 loop corrections, 
${\cal T}^{(+1)}_{corr}$, to the Born amplitude 
were discussed in Ref. \cite{pmmmk96}. 
We denote by ${\cal T}^{(+2)}_{corr}$
the $\nu=2$ loop and recoil corrections to the Born term. 
The recoil corrections to the Born term are reduced by 
$\left(m_\pi / m_N \right)^{\nu + 2}$.
The divergences contained in the graphs in Fig. 2 are 
canceled (``renormalized'')  
by counter-terms in ${\cal L}^{(2)}$ corresponding to
the five-point $(\pi N\!N\!N\!N)$ vertex diagram (Fig.6). 
The same terms renormalize a
part of the singularities in 
$M_{\mbox{\scriptsize IV}}$, Eq.(\ref{eq:3}), coming from Fig.3. 
The remaining singularities 
in $M_{\mbox{\scriptsize IV}}$ are similar to those in the graphs in Fig.4.
To eliminate the latter singularities, 
${\cal L}^{(2)}$ must contain further counter-terms 
of the pion-nucleon scattering vertex type \cite{fms98}.
This can be accomplished with the use of these counter-terms 
in graphs similar to the one in Fig.5.
%one should be able to generate contributions we need.
We let ${\cal T}_{corr}^{\prime (+2)}$ stand for
the $\nu$ = 2 transition operators that originate 
from such $\nu = 2$ counter-terms.
The complete set of the $\nu=2$ transition operators
includes ${\cal T}^{(+1)}_{corr}$, ${\cal T}^{(+2)}_{corr}$, 
and ${\cal T}^{\prime (+2)}_{corr}$, but
%However, 
we defer detailed discussion of these terms 
to a forthcoming paper \cite{dkmstwo} 
and concentrate here on the finite parts of the following
effective operators 
\footnote{Since we do not consider DW we leave out 
${\cal T}^{Born}$ }
\begin{equation}
{\cal T}\,= \, 
%{\cal T}^{{\rm Born}}
{\cal T}^{{\rm Resc}} 
+ {\cal T}^{(+2)} .  
\label{eq:calTtrunc}
\end{equation}

\paragraph{Numerical results and discussion}
The purpose of this section is purely illustrative: we wish to
have some idea as to the size of these corrections. 
A proper treatment of the derived transition operators 
must involve DW analyses, which we postpone to the forthcoming 
paper \cite{dkmstwo}, where DW modifications to our 
numerical estimates described below 
as well as changes due to the use of a  
non-static nucleon propagator will be discussed. 

To proceed, we must fix the free parameters 
of the previous expressions.
To NLO, as discussed in Ref. \cite{pmmmk96,cfmv96},
the three parameters,
$c_1$, $c_2$ and $c_3$ of Eq.(\ref{eq:kappakq}), 
enter into the pion rescattering operator ${\cal T}^{Resc}$. 
We shall use the three sets of parameters 
employed in Ref.\cite{slmk97}. 
Sets A, B and C in Table 1 summarize these values. 
Set A represents the central values of $c_1$, $c_2$ and $c_3$
determined in Ref.\cite{bkm95} using the experimental values of
the pion-nucleon $\sigma$ term,
the nucleon axial polarizability $\alpha_A $  
and the isospin-even s-wave
$\pi N$ scattering length $a^+$. 
Sets B and C represent typical ranges of allowed values 
%ambiguity that exist 
in the current determinations 
of theses parameters (see Ref.\cite{pmmmk96} for details). 

For simplicity we limit our consideration to the {\it threshold kinematics}, 
which means 
$q^\mu$ = $(m_\pi , \vec{0})$ and  
the single exchanged boson (pion) of Fig. 1(b) has 
the four-momentum $k^\mu$ = $(m_\pi/2, \vec{k})$ with $k^2$ = $- m_\pi m_N$. 
Then the final nucleon three momenta  are $\vec{p}^{\; \prime}_i $ = 0 
(for nucleon $i$ = 1, 2), 
and $\kappa (k, q)$ in Eq.(\ref{eq:kappakq})
is fixed at\cite{pmmmk96,cfmv96} 
\begin{equation} 
\kappa_{th} = \frac{m_\pi^2}{f_\pi^2} \left( 2 c_1 -\frac{1}{2} 
(c_2 + c_3 - \frac{g_A^2}{8 m_N}) \right)  . 
\end{equation} 
Quantities of interest here are the magnitudes of the  
finite parts of the $\nu$ = 2 transition operators, 
Eqs.(\ref{eq:2a})-(\ref{eq:4b}), 
relative to the magnitude of the pion rescattering operator,
${\cal T}^{Resc}$, Eq. (\ref{eq:Tplus1}). 
Let us denote these ratios 
\footnote{
This type of comparison is possible because, 
at threshold, effectively only one
kind of spin operator appears in the 
$pp\rightarrow pp\pi^0$ transition operator.} 
by $R_K \equiv M_{\mbox{\scriptsize K}}/{\cal T}^{Resc}$ 
(K = I, II, III, $\ldots$, VII).
Table 2 gives $R_K$'s for each of the parameter sets A, B, and C. 
We note that $R_I$, $R_{II}$ and $R_{III}$,
corresponding to the graphs in Fig. 2,
give quite substantial individual contributions 
but  $R_{II}$ and $R_{III}$ cancel each other
at threshold.\footnote{
With DW this cancellation will not occur due to the 
different energy and three-momentum dependences.}
%Especially $R_{II}$ and $R_{III}$, which at the threshold happen
%to have the same ratios, are large, ranging $3 \sim 7$ for the 
%three parameter sets studied.
Most remarkably, $R_{IV}$ corresponding to
the pion-pion rescattering diagram, Fig.3, 
is large, ranging $5\sim 10$.

The appearance of these large individual 
contributions calls for an explanation. 
The two-pion exchange diagrams in Figs. 2 and 3 all involve 
one-loop integrals with three or four propagators. 
These loop integrals,
for which typical four-momentum transfers $k$
are large,
can produce the factor $k^2 = - m_{\pi} m_{N}$
in the numerator multiplying the integral 
$I_0 ( \omega, v\cdot P , P^2)$ of eq.(\ref{e:io}). 
This factor turns out to be accompanied 
by some negative powers of $f_{\pi}^2$ coming from the vertices,
resulting in a large enhancement factor,
${\cal O}(k^2 f_{\pi}^{-2} = - m_{\pi} m_{N}  f_{\pi}^{-2} 
\simeq - 15)$.
This feature essentially explains the large 
size of the two-pion exchange diagrams,
although there are other diagram-specific 
numerical factors.

The two-pion exchange diagrams in Fig. 2 can perhaps be viewed 
as a part of an effective $\sigma$-meson exchange 
that Lee and Riska\cite{lr93} found to be important in 
$pp \rightarrow pp\pi^0$.
It has been shown via soft-pion arguments \cite{bd71}
that the effective $\sigma$-meson 
exchange can be understood as a two-pion exchange. 
The results in Ref.\cite{bd71} lead us to suspect that 
the few diagrams we consider here are insufficient 
to generate the full strength of isoscalar two-pion 
exchange between two nucleons, 
but our NNLO results are indicative 
of the importance of the two-pion exchange diagrams
for $pp \rightarrow pp\pi^0$. 
Similar HB$\chi$PT two-pion exchange diagrams have been  
considered in calculating the scattering amplitudes 
for higher partial waves in $NN$ collision\cite{kaiser98}. 
It is to be noted, however,
that the higher partial wave amplitudes,
which are only sensitive to peripheral NN scattering,
can probe two-pion exchange contributions
only for low three-momentum transfers. 
By contrast, 
in the $NN \rightarrow NN \pi$ reaction,
the two-pion exchange diagrams are probed 
in a very different kinematical regime 
of ``high" energy- and three-momentum transfers 
between the two nucleons. 
It is therefore not surprising 
that the roles of the two-pion exchange diagrams
in our calculation are very different 
from those discussed in Ref.\cite{kaiser98}.

The diagrams shown in Fig. 4 generate effective form factors
at the pion-nucleon rescattering vertex
in ${\cal T}^{(+1)}$. 
According to Table 2, 
the contributions of the corresponding operators, 
$R_V$ and $R_{VI}$, are 
less than 20\% and 15\% , respectively.
Thus, the higher chiral-order vertex corrections to ${\cal T}^{(+1)}$ 
are found to be small, as expected from the
general tenets of $\chi$PT.
Meanwhile, Table 2 shows that 
$R_{VII}\,=\,1.9  \sim 2.6$.
This means that the combined pion recattering term is 
given by the sum of Fig.1(b) and the Type VII contribution. 

Gedalin {\it et al.} \cite{gmr98} 
considered some NNLO diagrams within HB$\chi$PT.  
Numerically, they have found that the sum of 
$M_{\mbox{\scriptsize IV}}$ 
and $M_{\mbox{\scriptsize V}}$ 
(in our notation) is large. 
This feature is confirmed by our numerical result. 
It is worth emphasizing, however,
that of these two operators
$M_{\mbox{\scriptsize IV}}$ is by far the predominant one.
Although Gedalin {\it et al.} also considered
$M_{\mbox{\scriptsize II}}$ (our notation), 
they left out 
$M_{\mbox{\scriptsize I}}$ and $M_{\mbox{\scriptsize III}}$.
According to Table 2, 
$M_{\mbox{\scriptsize II}}$ and $M_{\mbox{\scriptsize III}}$
are of equal importance, and their individual contributions 
are comparable to that of $M_{\mbox{\scriptsize IV}}$.
Our final remark is 
that our results contain a new loop-integral 
$I_0(\omega,v\!\cdot\! p,p^2)$, Eq. (\ref{e:io}), 
which does not appear in Ref. \cite{gmr98}. 

\paragraph{Conclusions}
We have evaluated the $pp\rightarrow pp \pi^0$ transition 
operators to NNLO in chiral expansion.
It was found that $\chi$PT vertex corrections  to 
the lower chiral order transition operators,
${\cal T}^{Born}$ and  ${\cal T}^{Resc}$
are indeed small.
Thus, for this limited type of diagrams,
$\chi$PT expansion seems to be under control.
Meanwhile, the two-pion exchange contributions are 
found to be very large in our estimates.
This result is consistent with the expectation
that the $pp\rightarrow pp \pi^0$ reaction is sensitive 
to ``heavy"-meson exchanges between nucleons.  
The phenomenologically important $\sigma$-meson 
contributions \cite{lr93}
seem to have discernible ``representatives"
among the NNLO chiral perturbation diagrams considered here.
It is not obvious whether one can interpret 
the large contributions from 
the individual graphs in Figs. 2 and 3 
as evidence for the non-convergence of the $\chi$PT expansion. 
These types of graphs make their {\it first appearance}
only in the NNLO calculations,
and therefore the convergence question 
can only be settled by calculating corrections 
to these NNLO diagrams. 
We expect that the loop corrections to 
the individual  diagrams in Fig.2 
will be smaller in magnitude. However, two-pion 
exchange diagrams of chiral order $\nu$ = 3 might have 
magnitudes comparable to our $\nu$ = 2 terms 
since the diagrams in Fig.2 are only part of the 
effective $\sigma$- exchange.  

To simulate more realistic $\sigma$-exchange 
it may also be necessary 
to explicitly include intermediate $\Delta$-particles 
in Figs. 2 and 3, 
but that would require a thorough recalculation 
of many previous results.
In a forthcoming paper \cite{dkmstwo} 
we hope to present a detailed
discussion of DW calculations 
which are required to obtain 
realistic cross sections for $pp \rightarrow pp\pi^0$,
as well as the details of a renormalization procedure 
relevant to an NLLO calculation.
Due to large energy-momentum transfers 
involved in the $NN\rightarrow NN\pi$ reaction,
a full DW calculation can modify significantly
the numerical results reported in this Letter. 
Our finding that the ``recoil" corrections 
${\cal O}(m_N^{-1})$ to ${\cal T}^{Resc}$
are large points to the necessity of examining
the use of the static heavy-baryon ``propagator'',
$1/(v\!\cdot\! p)$.
As a first pragmatic step,
one can think of replacing $1/(v\!\cdot\! p)$ 
with $1/(v\!\cdot\! p - \vec{p}^{\; 2}/(2 m_N))$,
which implies, however, an extension of HB$\chi$PT adopted in this work.

We thank Shung-ichi Ando for pointing out an error 
in our calculation of the recoil correction to the 
pion rescattering diagram.
This work is supported in part
by the National Science Foundation,
Grant No. PHYS-9602000 
and by the Grant-in-Aid of
Scientific Research, the Ministry of Education,
Science and Culture, Japan, Contract No.07640405.

\begin{table}
\caption{The three sets of $c_{1,2,3}$ parameters, in units of GeV$^{-1}$, 
used in the text.}
\begin{tabular}{lllll} & {$~~~c_{1}$} & {$~c_{2}$} & {$~~~c_{3}$} \\
 \tableline
{A} & {$- 0.87$} & {$3.34$} & {$- 5.25$}  \\
{B} & {$- 0.87$} & {$4.5$} & {$- 5.25$}  \\
{C} & {$- 0.98$} & {$3.34$} & {$- 5.25$} \\
\end{tabular}
\end{table}

\begin{table}
\caption{
Sizes of the $K^{\rm th}$ type of diagrams, shown 
in Figs. 2 - 5, relative to the $\nu$ = 1 
pion rescattering diagram, Fig. 1(b). 
The ratio $R_K$ defined in the text is given for 
three sets (A, B and C) of parameters ($c_{1,2,3}$), 
see the text and Table 1.}
\begin{tabular}{lllll} & {$R_K^{\rm A}$} & {$R_K^{\rm B}$} & {$R_K^{\rm C}$} \\
 \tableline
{$K$ = I} & {$- 0.70$} & {$- 0.38$} & {$- 0.53$}  \\
{$K$ = II} & {$6.7$} & {$3.6$} & {$5.1$}  \\
{$K$ = III} & {$-6.7$} & {$-3.6$} & {$-5.1$}  \\
{$K$ = IV} & {$9.5$} & {$5.1$} & {$7.2$}  \\
{$K$ = V} & {$0.18$} & {$0.10$} & {$0.14$}  \\
{$K$ = VI}& {$0.14$} & {$0.08$} & {$0.11$}   \\ 
{$K$ = VII}& {$2.6$} & {$1.9$} & {$2.0$}   \\
\end{tabular}
\end{table}

%\newpage
\begin{figure}
\begin{center}
\epsfig{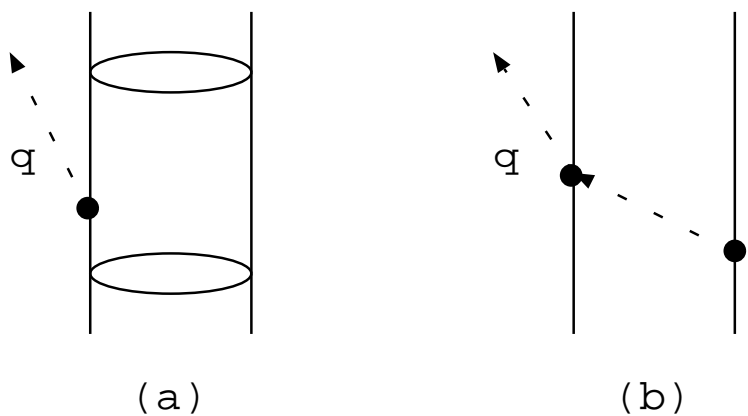} 
\end{center} 
\caption[]{Diagrams contributing to $pp\rightarrow pp\pi^0$:
Born term (a), the pion-rescattering term (b). The solid line denotes
a nucleon (proton), the dashed line a pion and the 
open ellipses  denote
initial-/final-state interactions.}
\end{figure}

%\newpage\section*{Figs. 2:} 
\begin{figure}
\begin{center}
%\vspace{2cm}
\epsfig{file=type1.eps,width=11cm} 

\vspace*{2cm}
\epsfig{file=type2.eps,width=7cm} 

\vspace*{2cm}
\epsfig{file=type3.eps,width=11cm}

\end{center} 
\caption[]{Two-pion exchange graphs of type I, II and III 
-the``cross-box" graphs.}
\end{figure}

%\newpage\section*{Fig.3:} 
\begin{figure}
\begin{center}
\epsfig{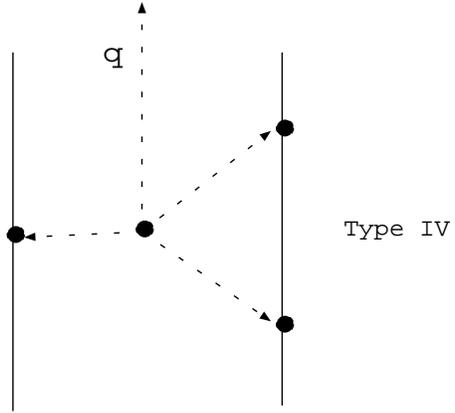} 
\end{center} 
\caption[]{Pion-pion s-wave rescattering graph of type IV.}
\end{figure}

%\newpage \section*{Fig.4: }
\begin{figure}
\begin{center}
%\vspace{2cm}
\epsfig{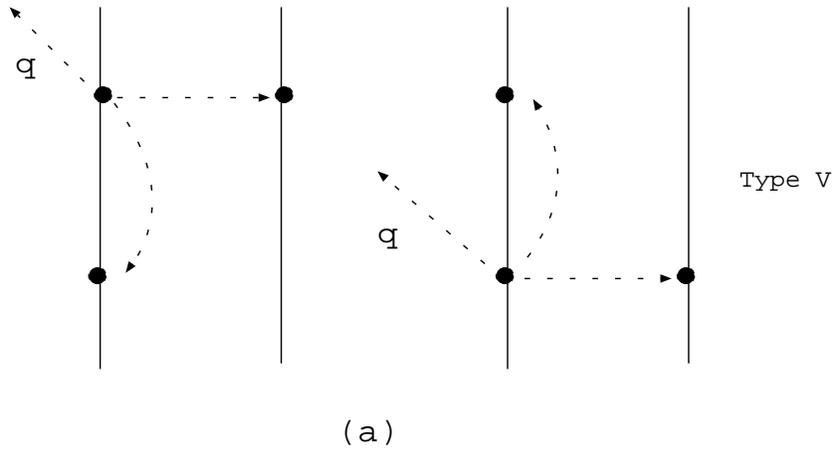} 

\vspace*{2cm}
\epsfig{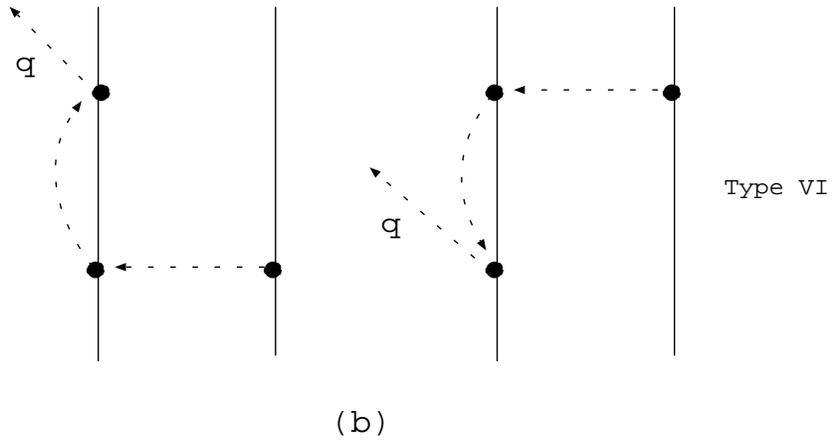} 
\end{center} 
\caption[]{Pion rescattering vertex corrections to Fig.1(b) of type V and VI.}
\end{figure}

%\section*{Fig6  }
\begin{figure}
\begin{center}
\epsfig{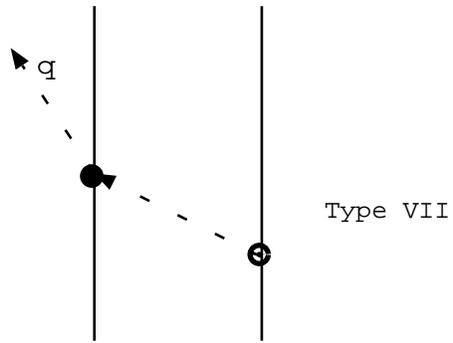} 
\end{center}
\caption[]{The pion-rescattering graph with $\bar{\nu}$ =2 
at rescattering vertex.}
\end{figure}

\begin{figure}
\begin{center}
\epsfig{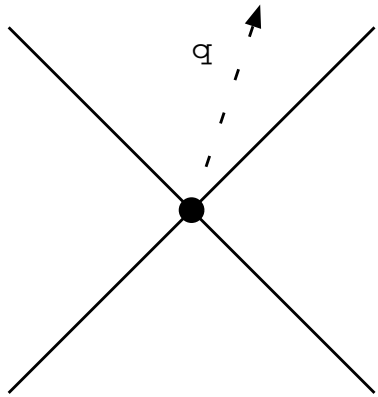}
\end{center}
\caption[]{The five-point contact-interaction counter-term graph.}
\end{figure}

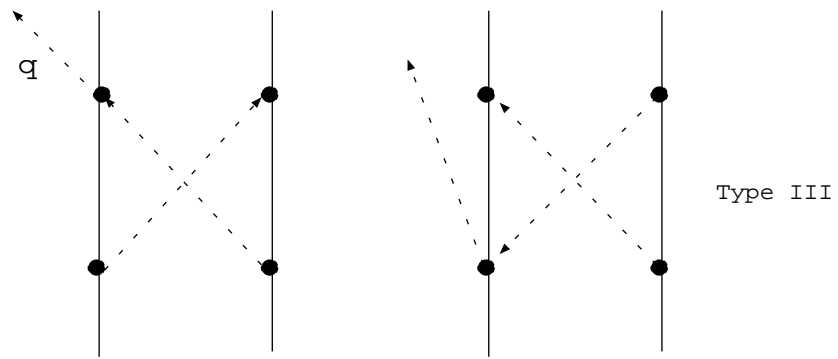
\end{document}